# Evidence lacking for a pending collapse of the Atlantic Meridional Overturning Circulation

Xianyao Chen[1,2] and Ka-Kit Tung[3,*]

[1] Frontier Science Center for Deep Ocean Multispheres and Earth System and Physical Oceanography Laboratory, Ocean University of China, Qingdao, China
[2.] Qingdao National Laboratory for Marine Science and Technology, Qingdao, China
[3] Department of Applied Mathematics, University of Washington, Seattle, WA, USA
[*] Corresponding Author: ktung@uw.edu

Matters Arising from Boers, *Nature Climate Change,* *11*, pages 680–688 (2021)

A catastrophic collapse of the Atlantic Meridional Overturning Circulation (AMOC) will have serious impacts on global climate. Based on eight AMOC proxies across the Atlantic basin *Boers*[1] claimed to have found Early Warning Signals (EWS) that point to its imminent collapse.  **Here we show (1) common to all eight of Boers' AMOC proxies are artificial rises in variance as observational coverage increased, giving false alarms; we demonstrate how climate models without EWS could be made to yield apparent EWS of a tipping point when sampled according to the increasing historical observational coverage. (2) Boers' salinity proxy was based on our work[2]; this and another of his subpolar AMOC proxy have their signs reversed erroneously.  The narrative that AMOC is declining towards collapse cannot be supported when the sign is corrected.**

The tipping point of a dynamical system has at least three possible types of EWS: increasing variance and autocorrelation[3], and slower recovery from perturbation as measured by the restoring rate $\lambda$. When the generalized least-squares regression is used in the calculation, the last type of EWS avoids giving false alarms when the increase is due to external noises having increasing variance and autocorrelation. However, all three EWS proposed by Boers[1] are vulnerable to another type of false alarms, that from increasingly denser data coverage, which affects "data", apart from external noise.

      At low coverage, the improving coverage is a more important contributor to the apparent increase of observed variance than real multidecadal changes.  Boers' algorithm for calculating EWS requires over 120 years of data, while observations were very sparse prior to 1950s. The dataset used was EN4, with $1°\text{x}1°$ resolution[4]. However, much of these grid boxes have no actual observation in early decades.  The ratio of boxes with at least one profile of observation during anytime in a month to the total number of boxes within a region is plotted (in red) in Figure 1a and b. Coverage is very low, around a few percent at this resolution for the decades before 1950, and never reaching 30% in the Southern Ocean, where coverage changes (both up and down) have the greatest impact on the changes in variance measured. When there is no observation, EN4's data defaults to

an annually periodic "climatology" without an anomaly, yielding a low, unchanging variance, as can be seen in the left panel of Fig.1. Occasional measurements appear as data "glitches". Even at subpolar Atlantic, the subsurface coverage never exceeded 50%.

In addition to being an EWS itself, increasing variance leads to the increasing restoring rate calculated by Boers using his generalized least-squares regression. When linearized around the current "equilibrium", Boers found that the perturbation is described by the linear system $\frac{d}{dt}x = \lambda x + \eta(t)$, where $\eta(t)$ is noise. Boers found AMOC to be currently stable as $\lambda$ is negative, but in successive 70-year sliding windows it is approaching 0 from below; the variance becomes higher because it is damped less and less. The year when the extrapolated $\lambda$ reaches 0 is the time of the "collapse", which was found to occur by 2020; see Figure 3 of Boers and Figure 1 below.

We find, however, that the positive slope of $\lambda$ shown by Boers—regarded as a reliable EWS—depends on the positive trend of variance rising from low value of early decades to the higher variance in the recent decades (see Figure 1). Importantly, the positive trend in the observed variance corresponds to the positive trend in observational coverage. $S_{NN2}$ is similar to $S_{NN1}$, and N similar to S.

We next show using climate models that it is the increasing observational coverage that causes the positive slope of $\lambda$ (see Figure 2). We choose two climate models, CCSM4 and GFDL-ESM2M, originally without EWS. They are regarded as the "truth" because models have complete coverage. Remarkably, when we subsample the model salinity proxies using the observational coverage, EWS appears, proving that these are false alarms because they differ from the truth.

In contrast, when we subsampled the infilled EN4 data, the apparent EWS found by Boers remains, meaning that EN4 behaves like its subsampled version so that the measured variance increases as the subsampled data include more and more real measurements—a false alarm. When 90% of the previous month's anomaly is retained in the gap-filling procedure called "EN4" by Boers, "memory" is enhanced but not the variance. Our conclusion on variance and restoring rate is not affected. See Figure 2, bottom row. EN4's infilled "data" at cells where there was no observation contain little additional information that could affect EWS.

The second part of our remarks deals with evidence provided by Boers[1] that AMOC is secularly declining. Because systematic in-situ measurements of AMOC's strength started only since 2004[5], proxies are used to detect the possible long-term trend. *Chen and Tung*[2] extended the record to 1950 by using a salinity-based proxy in the subsurface subpolar Atlantic ($S_{NN1}$), based on the physical consideration that a stronger (weaker) AMOC will transport more (less) salt to the subpolar North Atlantic. This was verified by us using the RAPID in situ measurements[5] since 2004 and satellite measured sea-surface height since 1993[6]. *Boers* claimed that his salinity proxy was based on *Chen and Tung* but we found that his two subpolar salinity proxies have an erroneous sign. The sign however does not affect the variance, which is the square of the anomaly, but it affects the author's narrative that AMOC is secularly declining. Using Boers' Figure 3b but correcting the sign error, the AMOC represented by $S_{NN1}$ and $S_{NN2}$ should have positive recent trends, with the positive trend of the latter being very pronounced. See Boers' Figure 3b but reverse the sign. The prediction of AMOC's collapse by 2020 should not be accompanied by decadal strengthening at the time of "collapse".

The South Atlantic salinity (S) from Zhu and Liu[7] highlighting the steep decline of AMOC starting from 1970 in Boers' Figure 3b may not be real. There was almost no subsurface observation in the Southern Atlantic prior to 1950s, leading to a dramatic increase in measured salinity later with the deployment of Argo floats[5]. Boers turned it into a negative trend by switching the sign of the index. This change in sign is justifiable: Southern Atlantic tends to have a salinity pile up when AMOC slows because this is the region from which AMOC is transporting salt away. The negative trend shown by Boers however was from an artificial positive trend caused by the rapid increase of observation. The same remark applies to the tropical ($S_N$) salinity index used by Boers, though to a lesser degree. There should not be a salinity pile up in subpolar latitudes as AMOC slows because subpolar Atlantic is where AMOC transports salty water to, not from.

The main SST proxy, $SST_{SG-GM}$, is based on the difference between the subpolar gyre SST and the global-mean SST[8]. It mainly reflects the negative of global warming, giving the narrative by construction that the AMOC is in a long-term decline as global warming increases in the industrial era. The observed cooling in the so-called "warming hole" of the subpolar gyre SST does not necessarily reflect the strength of the ocean circulation. He et al.[9] reproduced the warming hole and similar SST-based AMOC proxy variations under global warming in a slab ocean model, which does not even have an AMOC! Furthermore, because the global- (or hemispheric-) mean SST is used in the construction, they are greatly affected by the global (or hemispheric) coverage of observations, which is quite low until recently. Our previously expressed concern about observational coverage applies to these indices as well, regardless of how SST are preprocessed.

The original proposers[10] of such a SST-based proxy verified the proxy by comparing model AMOC with model proxy, using only one climate model, after dismissing 10 other models for not having realistic SST patterns. This verification may be subject to selection bias. Using a large collection of climate models we compare model AMOC with model SST-based proxy of the AMOC using correlation coefficients and find that the agreement is not statistically significant. Whether the modeled SST compares favorably with the observed SST is irrelevant in this comparison of model truth vs model proxy. Here we extract the information from Boers' existing figures. His Figure 6c lists the correlation coefficients between model truth and model proxy. In one model the proxy has no correlation with the truth and the other a negative one. Only in 3 out of 15 models do the correlation coefficient reach 0.65, with the rest clustering around 0.3. Even 0.65 may not be high enough given the sensitivity of Boers' algorithm to slight differences in input time series. In fact, Boers found that the existence of EWS in climate models when using the true AMOC strength differs from that for the proxies.

Boers defended his use of *Caesar et al.*'s SST-based proxy[8] by implying that it compared favorably with *Chen and Tung*[2]. In that comparison since 1950, both AMOC proxies have no secular trend, but have large (3 Sv) up and down multidecadal variations, likely natural. The 15% (3 Sv) decline of AMOC claimed by Caesar et al., which was the basis of Boers' own claim of "weakening of AMOC over the past 150 years", was from an AMOC peak in 1950 and a subsequent trough one and half cycles later, whereas there is no trend if measured correctly from peak to peak or from trough to trough.

Boers' results are self-contradictory: one proxy shows EWS exists while another does not in the same model. One observed proxy indicates AMOC is declining while another increasing. They are certainly not "spatially coherent". These proxies are supposed to represent the same single AMOC.

**Method**

We use the same code provided by Boers[1] to calculate $\lambda$ as the regression coefficient of $\frac{dx}{dt}$ in a generalized least squares regression against *x*. Sliding 60-year window is used. Instantaneous variance is calculated as the square of the deviation of the time series from its nonlinear secular trend[11]. A "glitch" in S in May 1949 is removed before the calculation of the instantaneous variance in Figure 1. In "subsampling", only data in gridpoints with observation are averaged to obtain the proxy. "EN4 filling" was introduced in Boers' Reply, where missing data are filled with 90% of previous month's data.

**The authors declare no competing interests.**

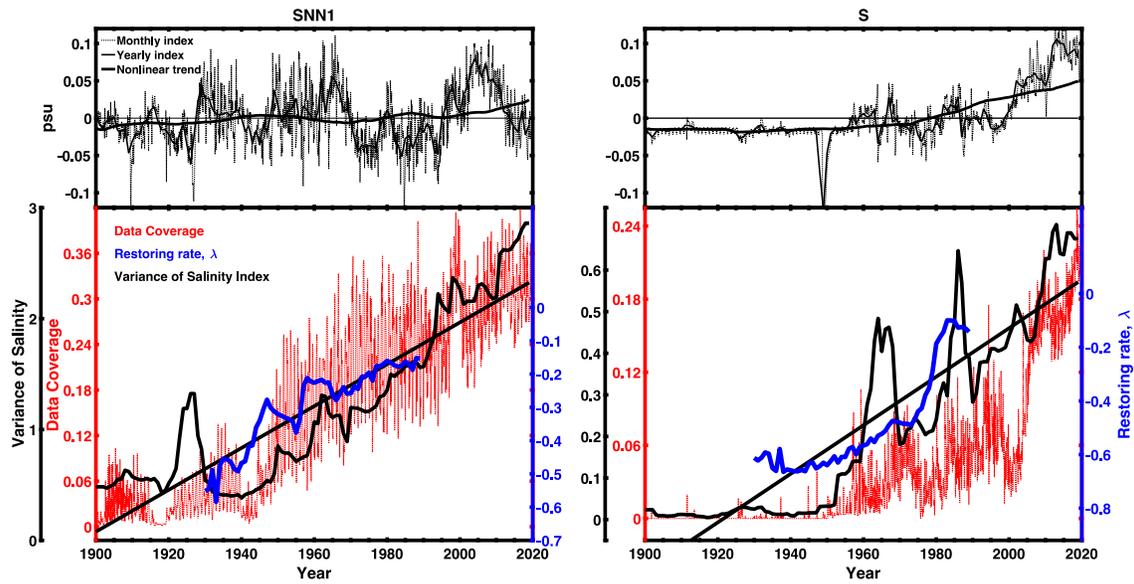

**Figure 1 Apparent Early Warning Signals arise as observational coverage improves**: (left) Subpolar salinity ($S_{NN1}$), and (right) Southern Atlantic salinity (S), 0-300m, from EN4. Top: monthly salinity (dashed line), yearly (thin solid line), and 50-year running mean (thick solid line). Bottom: observational coverage ratio (dashed red), restoring rate $\lambda$ calculated using the same generalized least-squares regression as Boers[1](blue), and instantaneous variance (black) calculated from annual mean data, with its least-squares linear trend indicated by a black straight line.

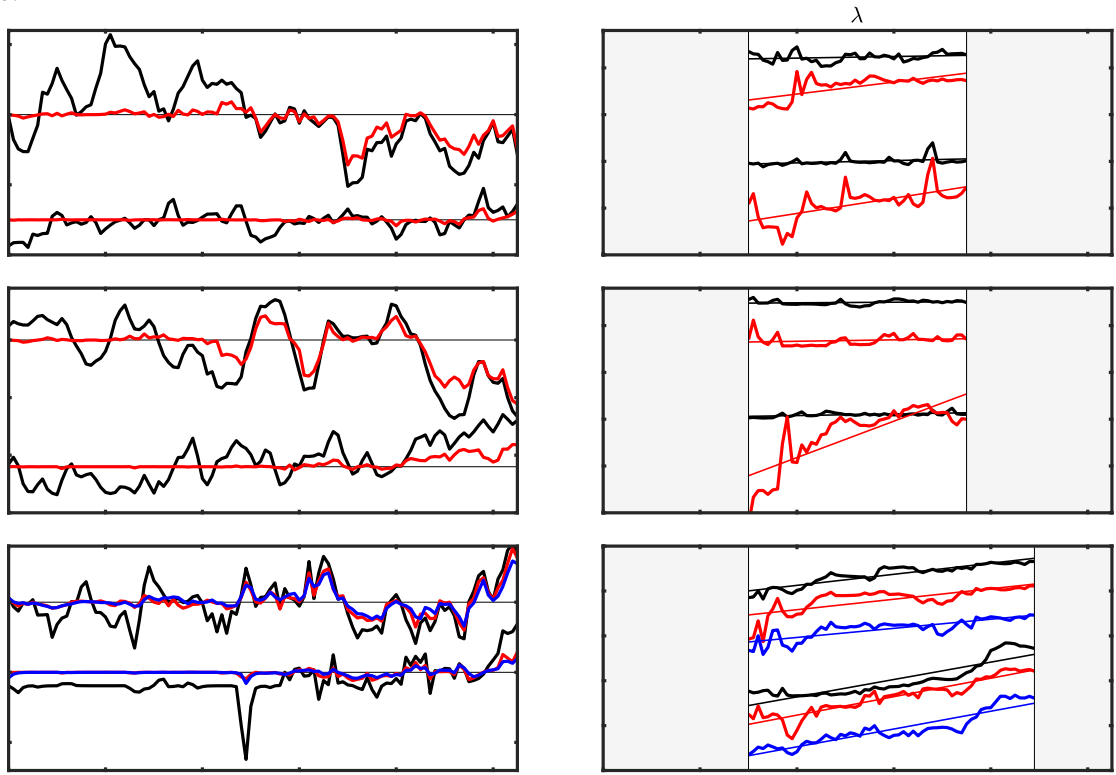
5

**Figure 2. Subsampling model and EN4 data**: Salinity anomaly averaged over the Subpolar and Southern Atlantic regions (left); the restoring rate and its linear trend (right). The black curves are from the original model output, the red curves from their subsampled version. Bottom row: Black curves are from the EN4 dataset, red curves from its subsampled version; the blue curves are for "EN4-filling". The "EN 4-filling" gives practically the same EWS as our subsampled data, when correctly applied only to the spatial points historically without observation, unlike Boers' unrealistic synthetic data where the temporal filling is applied to data gaps "artificially introduced", without distinguishing those spatial points with observation from those without, which have different statistics.